\documentclass{iau} 
\usepackage{graphicx}
\usepackage{enumitem}

\title[Dynamical Regularities in Galaxies] 
{Dynamical Regularities in Rotating Galaxies}

\author[McGaugh, Lelli, Li, \& Schombert]   
{Stacy McGaugh$^1$, Federico Lelli$^2$, Pengfei Li$^1$
 \and Jim Schombert$^3$}

\affiliation{$^1$Department of Astronomy, Case Western Reserve University, Cleveland, OH 44106, USA \\ 
$^2$European Southern Observatory, 85748, Garching bei M\"unchen, Germany  \\
$^3$Department of Physics, University of Oregon, Eugene, OR 97403, USA
}

\pubyear{2019}
\volume{353}  
\jname{Galactic Dynamics in the Era of Large Surveys}
\editors{M.\ Valluri \& J.\ Sellwood}
\begin{document}

\maketitle

\begin{abstract}
Galaxies are observed to obey a strict set of dynamical scaling relations.
We review these relations for rotationally supported disk galaxies spanning many decades in mass, surface brightness, and gas content.
The behavior of these widely varied systems can be summarized with a handful of empirical laws connected by a common acceleration scale. 
\end{abstract}


The dynamics of disk galaxies are well-organized, following a variety of well-established scaling relations.
These relations are remarkably tight, 
and can be summarized with a few simple rules: \\

\begin{enumerate}[label=\arabic*.\  ]

\item \textbf{Flat Rotation Curves}  

\begin{itemize}

\item[] The rotation curves of galaxies tend towards an approximately constant rotation speed that persists to indefinitely 
large radii \cite[(Rubin et al.\ 1978, 1980, Bosma 1981a,b)]{Rubin1978,Rubin1980,Bosma1981a,Bosma1981b}.
\end{itemize}

\item \textbf{Renzo's Rule} 

\begin{itemize}
\item[] For any feature in the luminosity profile there is a corresponding feature in the rotation curve, and vice versa \cite[(Sancisi 2004)]{renzorule}. 
\end{itemize}

\item \textbf{The Baryonic Tully-Fisher Relation (BTFR)}

\begin{itemize}
\item[] The amplitude of the flat rotation speed of a galaxy correlates with its baryonic mass (the sum of stars and gas:
\cite[McGaugh et al.\ 2000, Lelli et al.\ 2016a, 2019]{btforig,sparcbtfr,Lelli2019}).
\end{itemize}

\item \textbf{The Central Density Relation (CDR)}

\begin{itemize}
\item[] The dynamically measured central mass surface density of a galaxy correlates with its 
photometrically measured central surface brightness \cite[(Lelli et al.\ 2013, 2016c)]{Lelli2013,sparc_cdr}.
\end{itemize}

\item \textbf{The Radial Acceleration Relation (RAR)} 

\begin{itemize}
\item[] The observed centripetal acceleration correlates with that predicted by the distribution of 
baryonic mass \cite[(McGaugh et al.\ 2016, Lelli et al.\ 2017, Li et al.\ 2018)]{RAR,OneLaw,LiRARfit}. \\

\end{itemize}

\end{enumerate}

These rules have been established over the years by the efforts of many astronomers working across optical, infrared, and radio wavelengths.
For brevity, we illustrate these rules utilizing recent work for which the mass distributions of both stars and gas are well constrained:
the SPARC (Spitzer Photometry and Accurate Rotation Curves) database \cite[(Lelli et al.\ 2016b)]{SPARC} supplemented by the 
gas-rich galaxies discussed by \cite[McGaugh (2012)]{M12}. 
The SPARC database contains galaxies for which both HI data cubes and Spitzer [3.6] surface photometry are available, providing
a comprehensive picture of the distribution of both stars and gas. The supplementary gas rich galaxies lack Spitzer photometry,
but are so gas dominated that optical data suffice to trace the minority stars. 




The ideal galaxy sample includes all galaxies within a suitably large volume of the universe \cite[(e.g., Eckert et al.\ 2015)]{RESOLVE}.
In practice, this can never be achieved: there is always a minimum luminosity and surface brightness below which galaxies cannot be detected.
To make matters worse, the vast majority of galaxy catalogs are magnitude limited rather than volume limited. This strongly biases samples
to the brightest objects that exist in the intrinsic distribution \cite[(McGaugh et al.\ 1995)]{MBS1995}. 

\begin{figure}[t]
\begin{center}
 \includegraphics[width=5.4in]{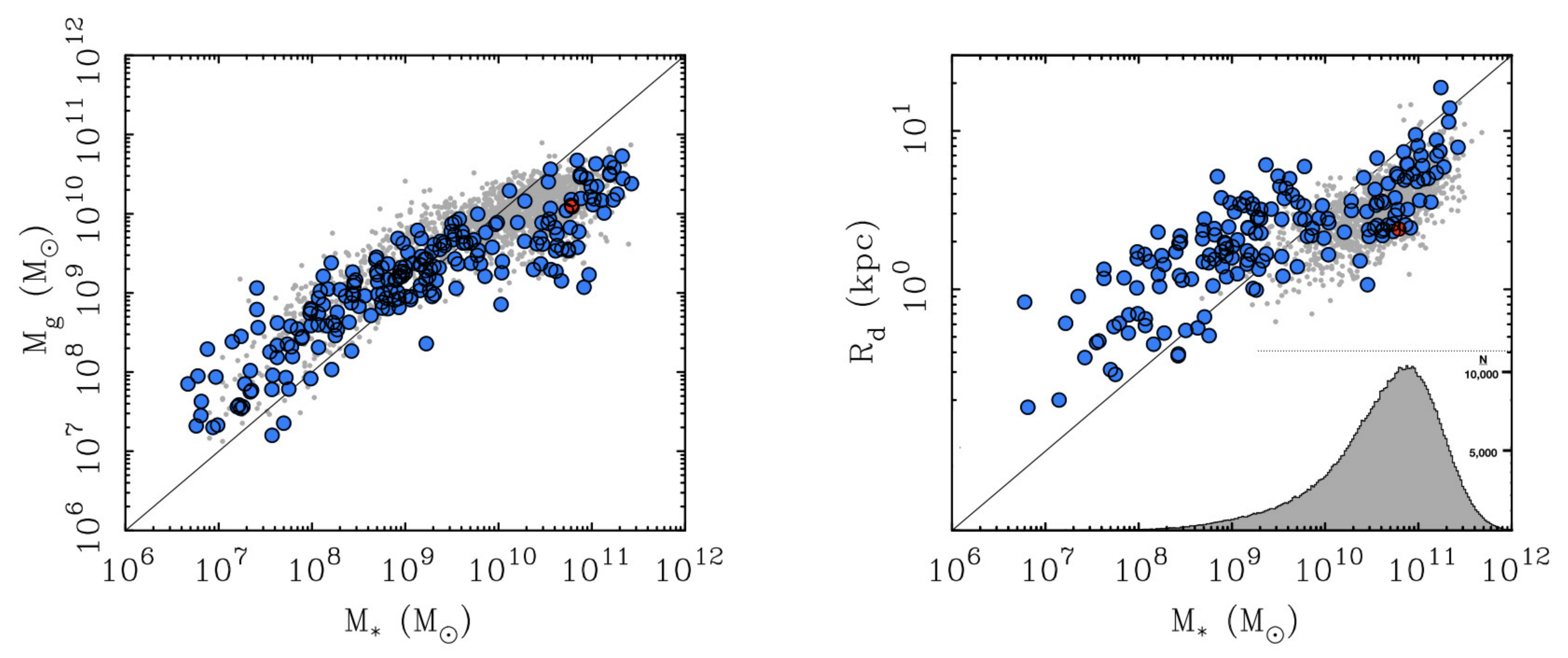} 
 \caption{The stellar masses of rotating galaxies compared to their gas masses (left) and disk scale lengths (right). 
 Blue points are galaxies in the SPARC database \cite[(Lelli et al.\ 2016b)]{SPARC} and the gas rich galaxies discussed by \cite[McGaugh (2012)]{M12}.
 The location of the Milky Way is noted in red \cite[(McGaugh 2016)]{M16}: it is a typical bright spiral.
  Grey points at left are the SDSS sample of \cite[Bradford et al.\ (2015)]{Bradford2015}; at right that of \cite[Courteau et al.\ (2007)]{C2007}.
  The line at left is the line of equality ($M_* = M_g$); that at right is a line of constant surface brightness ($M_* \sim R_d^2$).
 The inset at lower right shows the raw number of galaxies in SDSS DR7 as a function of stellar mass.
   \label{MstMg}}
\end{center}
\end{figure}

It being impossible to obtain a perfect galaxy sample, the next best thing is to sample randomly across all
decades of the mass function. The mass function rises to lower masses \cite[(Moffett et al.\ 2016)]{GAMA}, 
so a representative sample will have more low mass than high mass galaxies.
This is the opposite of what happens in magnitude-limited samples, where the numbers of low mass galaxies decline
sharply below a peak around $M_* = 5 \times 10^{10}\;\mathrm{M}_{\odot}$. While an obvious statement, Fig.\ \ref{MstMg} makes viscerally apparent
how stark the difference is between declining apparent numbers and increasing intrinsic numbers of low mass galaxies.
The sample discussed here provides a much broader perspective in terms of mass, surface brightness, and gas content
than is available from the magnitude limited samples that pervade the literature.


\begin{figure}[t]
\begin{center}
 \includegraphics[width=5.4in]{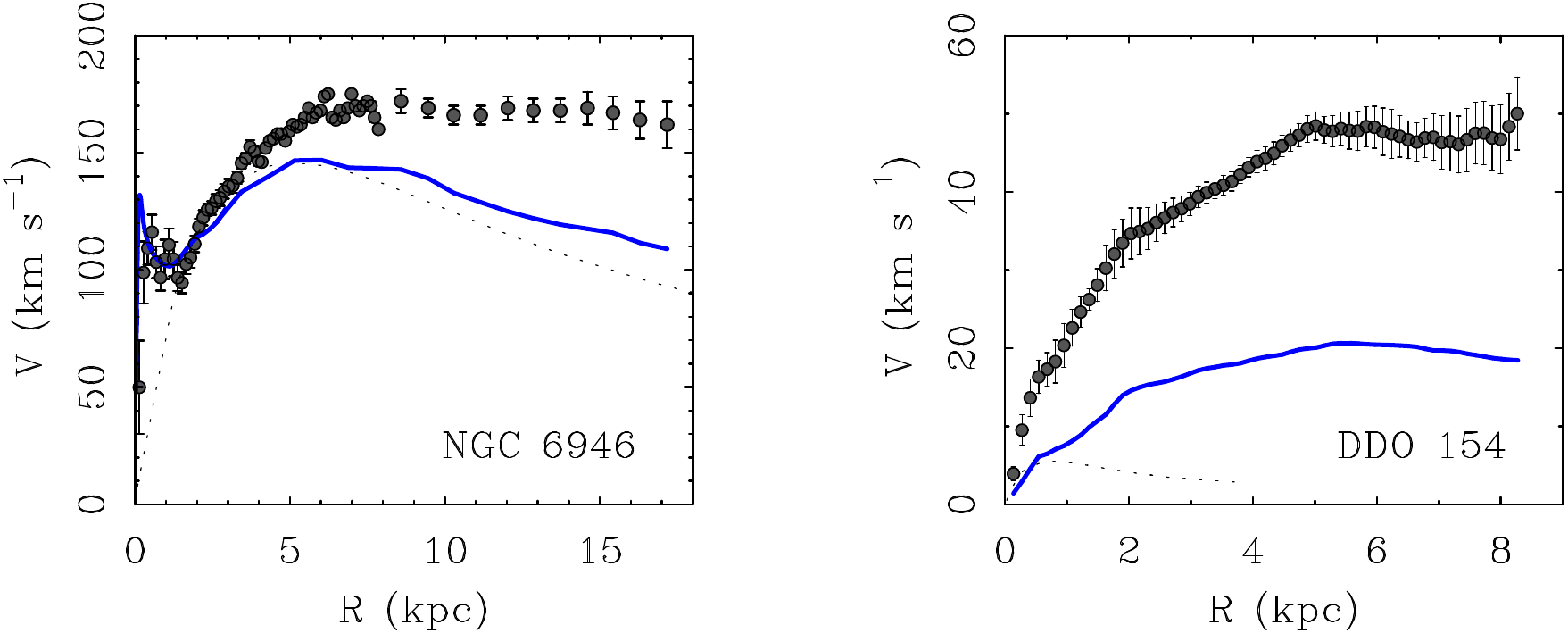} 
 \caption{Rotation curves (points) and mass models (lines) for the spiral galaxy 
 NGC 6946 \cite[(left: Blais-Ouellette et al.\ 2004, Daigle et al.\ 2006, Boomsma 2007)]{BO2004,D2006,Boomsma2007}
 and the gas rich dwarf DDO 154 \cite[(right: de Blok et al.\ 2008)]{THINGS}.
 In neither case is an exponential disk (dotted line) an adequate approximation for the mass model:
 proper treatment requires numerical solution of the Poisson equation.
   \label{renzorule}}
\end{center}
\end{figure}

\begin{figure}
\begin{center}
 \includegraphics[width=3.4in]{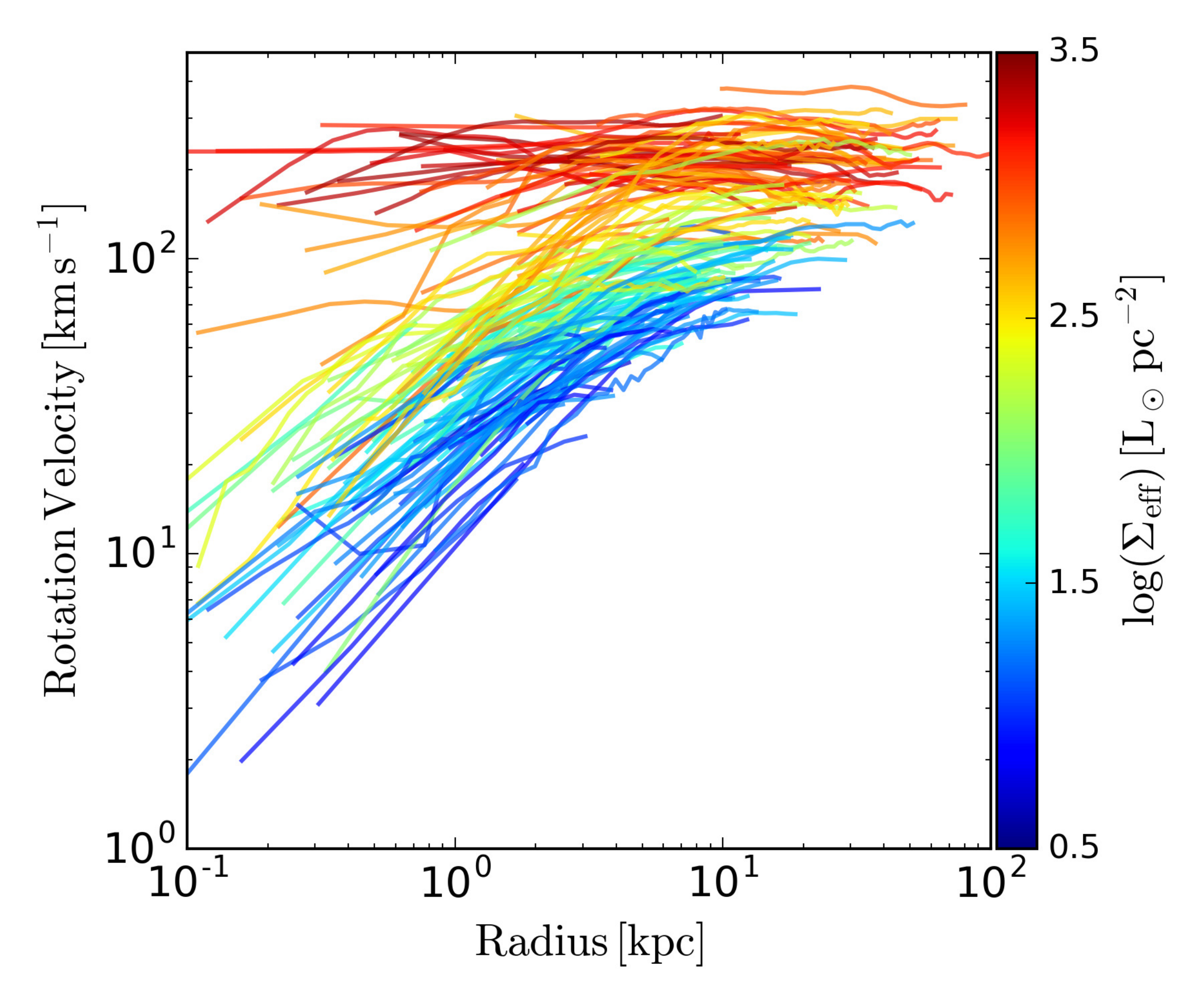} 
 \caption{Rotation curves of galaxies in the SPARC database \cite[(Lelli et al.\ 2016b)]{SPARC}
 color coded by 3.6 micron effective surface brightness. Note the 
 near-perfect rainbow sequence from the slowly rising
 rotation curves of LSB galaxies (blue) to the steeply rising ones of HSB galaxies (red).
 The dynamics of galaxies is closely connected to the distribution of light.
   \label{RCs}}
\end{center}
\end{figure}

\noindent\textbf{1.\ Flat Rotation Curves} are a sufficiently famous result that they require little review. Examples can be seen in 
Figures \ref{renzorule} and \ref{RCs}. That rotation curves become flat is, at this juncture, a \textit{de facto} Law of Nature akin to Kepler's Laws.

Of course, no rotation curve is \textit{perfectly} flat in the sense that $dV/dR = 0.000$. Rather, there is inevitably a range of radii over which
the rotation speed is nearly constant, within 5\% \cite[(Lelli et al.\ 2016a)]{sparcbtfr}. More generally, there is a tendency for the rotation
curves of bright galaxies to rise steeply then fall a little before flattening out \cite[(Noordermeer et al.\ 2007)]{Noord2007}, 
while those of faint galaxies tend to rise gradually, rolling over
toward flatness only slowly \cite[(de Blok \& McGaugh 1996, Swaters et al.\ 2009)]{dBM96,}. Often times, data do not reach far enough out to clearly see this: 
the data taper off while $V(R)$ is still rising \cite[(Stark et al.\ 2009, Trachternach et al.\ 2009)]{Stark,Trach}.
However, in those cases where more extended data have been obtained, the flattening has always been seen \cite[(de Blok et al.\ 2008)]{THINGS}.

\noindent\textbf{2.\ Renzo's Rule} highlights the correspondence between detailed features measured independently in the kinematics
and photometry of galaxies. This is natural when stars dominate the mass:
features in the dominant mass distribution must also be reflected in the gravitational potential. 
Renzo's rule also applies in low surface brightness (LSB)
galaxies, where the correspondence persists despite the mass discrepancy being large. 
This is \textit{not} natural, as a dynamically hot, quasi-spherical dark matter halo cannot support the same features as a dynamically cold,
thin baryonic disk \cite[(Binney \& Tremaine 1987)]{BT87}.



Fig.\ \ref{renzorule} illustrates Renzo's rule in both high and low surface brightness galaxies.
The inner shape of the mass model and rotation curve in NGC 6946 is driven by a compact bulge component
that contains a mere 6\% of the total light. Similarly, the mass model of DDO 154, which is dominated by gas 
($\mathrm{M}_{\mathrm{HI}}/\mathrm{L}_{[3.6]} = 5.2$, implying $f_g \approx 0.93$), 
has kinks around $R = 0.5$, 2, and 5 kpc that are also apparent in the rotation curve.
This occurs despite the baryons being sub-dominant at essentially all radii.
The general correspondence between surface brightness and kinematics is also apparent in Fig.\ \ref{RCs}.

\noindent\textbf{3.\ The Baryonic Tully-Fisher Relation} is a generalization of the original \cite[Tully \& Fisher (1977)]{TF77} relation to include gas mass
as well as stellar luminosity. Bright galaxies are star dominated (Fig.\ \ref{MstMg}), so a tight correlation between luminosity and 
rotation speed give a Tully-Fisher relation for galaxies with $M_* > 10^9\;\mathrm{M}_{\odot}$. Below this mass scale, 
the relation deteriorates. This is simply because an important mass component --- the gas --- has been neglected. A continuous
relation is restored when both stars and gas are included. The physics underpinning the relation is concerned with the total amount of
baryonic mass, but not whether it is stars or gas. 

The BTFR is a simple power law of the form $M_b \propto V^x$.
Various measures for the rotation speed give similar but subtly different Tully-Fisher relations, differing in details like the slope $x$ and scatter.
To the best we are able to determine, the scatter is minimized when the flat rotation velocity can be measured. 
This cannot be said of other measures like the linewidth \cite[(Lelli et al.\ 2019)]{Lelli2019}. Indeed, the intrinsic scatter 
of the $V_f$ BTFR --- that left over after accounting for scatter caused by observational uncertainties --- 
is remarkably small for an extragalactic correlation, $\sim 0.1$ dex \cite[(Lelli et al.\ 2016a)]{sparcbtfr}. 
An irreducible component of the intrinsic scatter is that in stellar mass-to-light ratios ($M_*/L$). 
At near-IR wavelengths, stellar population models anticipate a scatter
from variations in the star formation history to be 0.1 to 0.15 dex \cite[(Bell \& de Jong 2001, Portinari et al.\ 2004, Meidt et al.\ 2014)]{BdJ01,P04,Meidt}.
This consumes the entire budget for intrinsic scatter in the BTFR, leaving essentially no room for variation in the galaxy-averaged IMF or 
scatter in the halo mass-concentration relation. 

The scatter in the BTFR has declined steadily as the data have improved.
Every time a new type of galaxy is identified and measured, it falls
close to the BTFR but not necessarily right on. As improved observations are made, the discrepant cases become more consistent with the BTFR.
A failure to measure $V_f$ combined with the inevitable systematic uncertainties in distances and inclinations 
above and beyond those indicated by the random errors in these quantities are the most common problems.
Experience leads us to anticipate a similar evolution from outliers
to adherents as new galaxies are measured on the fringes of current knowledge.

\begin{figure}[t]
\begin{center}
 \includegraphics[width=3.4in]{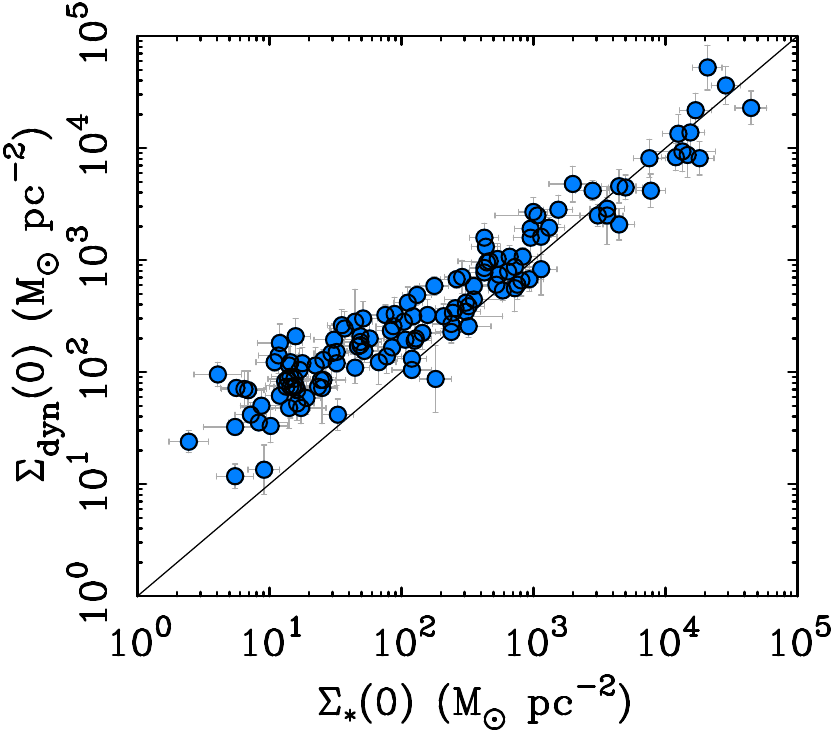} 
 \caption{The central density relation \cite[(Lelli et al.\ 2016c)]{sparc_cdr}.
 The dynamical mass surface density implied by the rate of rise of the rotation curve correlates with the stellar surface brightness
 at the centers of galaxies. Care is taken to measure both quantities in the same region as $R \rightarrow 0$ subject to the limitations of resolution. 
 The data follow the 1:1 line at high surface densities, where stars dominate the mass budget. 
 The data peel away from the 1:1 line at low surface brightness, evincing a mass discrepancy even at the centers of LSB galaxies.
   \label{CDR}}
\end{center}
\end{figure}

\noindent\textbf{4.\ The Central Density Relation} is a relation between the central surface brightness of galaxies and their dynamical mass surface density
(Fig.\ \ref{CDR}). This would be a trivial statement if there were no mass discrepancy. 
Indeed, at high surface brightness (HSB), there is a 1:1 relation between
surface brightness and mass surface density, as expected when stars dominate. However, as the surface brightness declines, the stars no longer suffice
to account for the mass budget, and the data depart from the 1:1 line. Despite the increasing need for dark matter towards the centers of LSB galaxies,
the correlation persists. The dynamical surface density can be predicted from the surface brightness of the sub-dominant stars.
Similar result can be seen in \cite[de Blok \& McGaugh (1996)]{dBM96}, \cite[McGaugh \& de Blok (1998)]{MdB98a},
\cite[Swaters et al.\ (2012, 2014)]{S2012,S2014} and \cite[Lelli et al.\ (2013)]{Lelli2013}. 

\begin{figure}[t]
\begin{center}
 \includegraphics[width=5.4in]{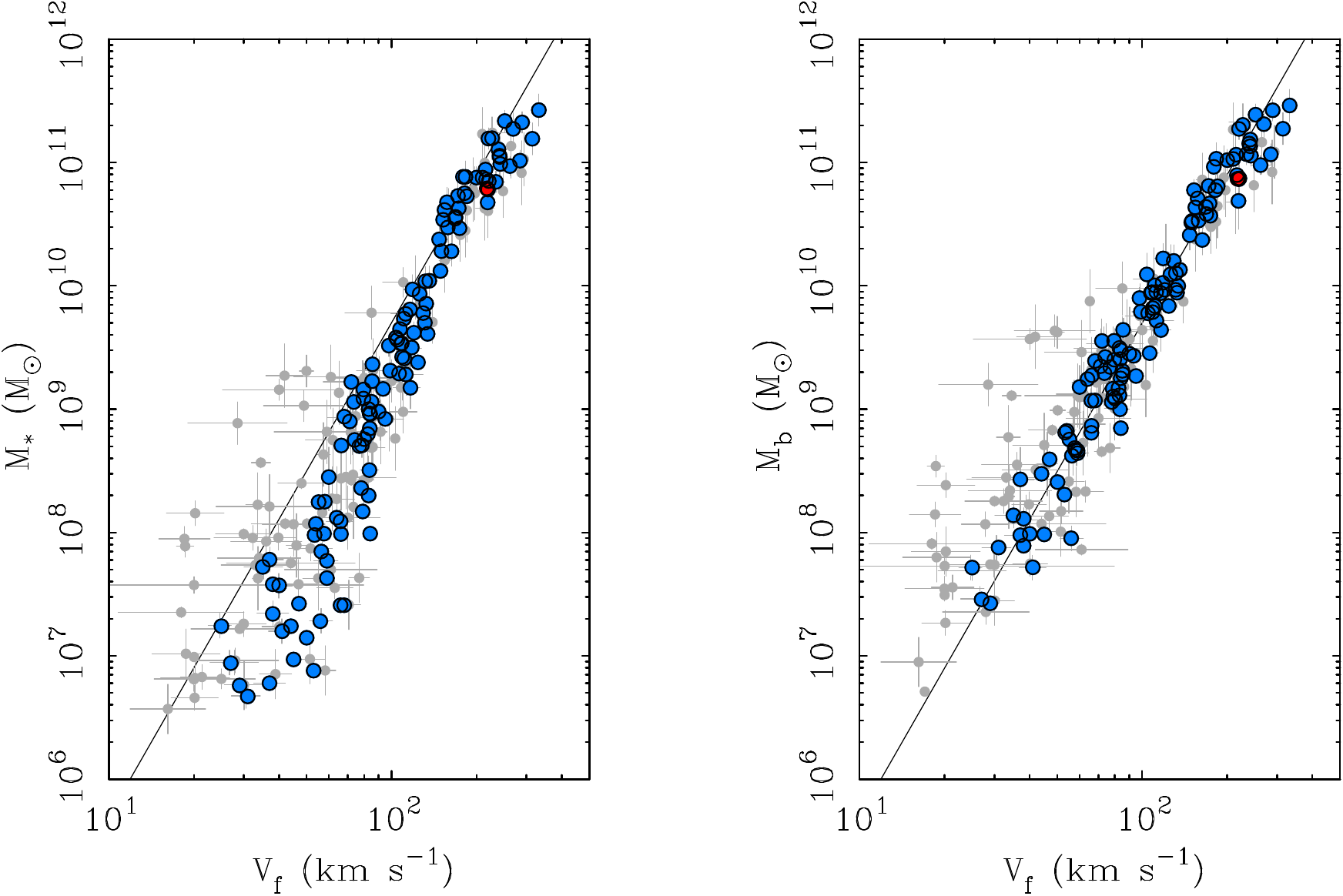} 
 \caption{The stellar mass (left) and baryonic Tully-Fisher relation (right). Baryonic mass (stars plus gas) correlates strongly with the flat rotation speed.
 Data from \cite[Lelli et al.\ (2016b)]{SPARC} and \cite[McGaugh (2012)]{M12} are shown as
 blue points if both axes are measured with at least 20\% accuracy; less accurate data are shown in grey.
 The latter include cases for which the rotation curve does not extend far enough to measure $V_f$, in which case
 the last measure point is used. These cases are systematically offset to lower velocity.
 The quantity $\mathrm{g}_{\mathrm{TF}} = \chi V_f^4/(G M_b)$ defines a line of constant acceleration, 
 illustrated here for $\mathrm{g}_{\mathrm{TF}} = 1.2\;\times10^{10}\;\mathrm{m}\,\mathrm{s}^{-2}$
 with $\chi = 0.8$ to account for the cylindrical geometry of disks.
 The location of the Milky Way is noted in red.
    \label{BTFR}}
\end{center}
\end{figure}

The CDR can be seen by inspection in Fig.\ \ref{RCs}, where the color coding by surface brightness results in a near-perfect rainbow.
Low surface brightness galaxies have slowly rising rotation curves, high surface brightness galaxies have rapidly rising ones.
The dynamics responds to the distribution of luminous mass.

The CDR contrasts with the BTFR in that the CDR depends on the distribution of luminous mass as quantified by the surface brightness,
while the BTFR depends only on the total baryonic mass and not its distribution. Indeed, one can find examples of galaxies of the same mass
but different surface brightness \cite[(de Blok \& McGaugh 1996, Tully \& Verheijen 1997)]{dBM96,TV97}. 
Such pairs of galaxies are indistinguishable in the BTFR, but reside in different locations on the CDR.
This has more recently come to be called `diversity' \cite[(Oman et al.\ 2015)]{Oman}, in which the rotation velocity measured at
small radius ($R = 2$ kpc) differs for galaxies of similar mass. This variation is entirely accounted for by the CDR: LSB galaxies have more slowly rising
rotation curves than HSB galaxies, even those of the same mass, so $V(R=2\;\mathrm{kpc})$ differs even when $V_f$ is the same.

\begin{figure}[t] 
\begin{center}
 \includegraphics[width=5.4in]{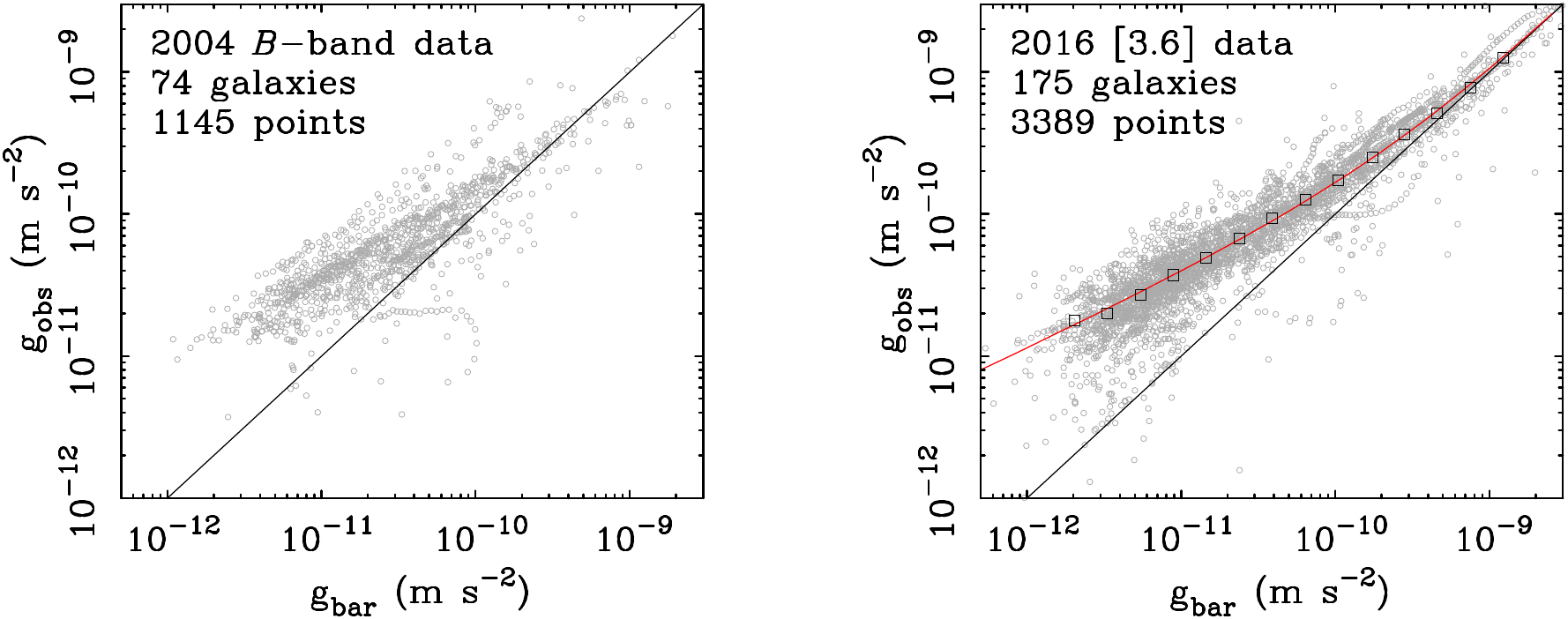} 
 \caption{The radial acceleration relation. The centripetal acceleration observed in the rotation curve,
 $\mathrm{g}_{\mathrm{obs}} = V^2/R$, correlates with that predicted by the observed distribution of baryons, 
 $\mathrm{g}_{\mathrm{bar}} = -\partial \Phi_{\mathrm{bar}}/\partial R$, 
 obtained from numerical solution of the Poisson equation applied to the observed distribution of stars and gas.  
 The data available to \cite[McGaugh (2004)]{M04} are shown in the left panel; those from \cite[McGaugh et al.\ (2016)]{RAR} in
 the right panel. All available data are shown with a population synthesis $M_*/L$ and 
 without selection for quality control. A few individual galaxies stand out in the left-hand panel, presumably because the 
 $B$-band $M_*/L$ is not always correctly predicted by population synthesis models. Nevertheless,
 the main relation was already clear twenty years ago \cite[(McGaugh 1999)]{M99}.
 Individual galaxies do not distinguish themselves in the right panel when Spitzer [3.6] data are used to trace the stellar mass distribution
 \cite[(Lelli et al.\ 2017)]{OneLaw}. Open squares show the data binned;
 the trend is well fit by a simple function ${\cal F}(\mathrm{g}_{\mathrm{bar}})$ (red line --- see text). 
   \label{RARfig}}
\end{center}
\end{figure}

\textbf{5.\ The Radial Acceleration Relation} is a correlation between the observed centripetal acceleration ($\mathrm{g}_{\mathrm{obs}} = V^2/R$)
and that predicted by the observed distribution of baryons ($\mathrm{g}_{\mathrm{bar}} = - \partial \Phi_{\mathrm{bar}}/\partial R$).
These two quantities are measured independently, and should correspond in a 1:1 fashion in a universe without dark matter.
This is true at high acceleration (Fig.\ \ref{RARfig}), which corresponds to high surface brightness through the Poisson equation.
As the acceleration declines, the data peel away from the 1:1 line in the same fashion as seen in the CDR.
Low surface brightness means low acceleration means a large mass discrepancy.

\cite[Sanders (1990)]{S90} identified an empirical correlation between the amplitude of the mass discrepancy and acceleration at
the last measured point along then-available rotation curves. This was generalized to include every resolved point by
\cite[McGaugh (1999)]{M99}. The relation has steadily improved as more data have accumulated.
Fig.\ \ref{RARfig} shows the progress from the data available to \cite[McGaugh (2004)]{M04} to now \cite[(McGaugh et al.\ 2016)]{RAR}.
In both cases, the axes are measured independently using a stellar population estimator for converting starlight to stellar mass.
A tremendous step forward has been provided by deep, near-IR data from \textit{Spitzer} that
provide an excellent tracer of stellar mass for computing the baryonic gravitational potential $\Phi_{\mathrm{bar}}$. 
The simplest possible assumption that all galaxies have the same [3.6] mass-to-light ratio 
\cite[(Schombert et al.\ 2014, 2019)]{SM14b,SML19} suffices to construct a relation in which individual galaxies
do not stand out. This contrasts with the situation for optical data, which always has a few outliers due to misestimates of $M_*/L$.

The trend of the data in Fig.\ \ref{RARfig} can be described by a simple function ${\cal F}(\mathrm{g}_{\mathrm{bar}})$
with a single characteristic scale, $\mathrm{g}_{\mathrm{\dagger}}$:
\begin{displaymath}
\mathrm{g}_{\mathrm{obs}} = {\cal F}(\mathrm{g}_{\mathrm{bar}})
 = \frac{\mathrm{g}_{\mathrm{bar}}}{1-e^{-\sqrt{\mathrm{g}_{\mathrm{bar}}/\mathrm{g}_{\mathrm{\dagger}}}}}.
\label{eq:RFRfit}
\end{displaymath}
The detailed shape of the baryonic mass model of each galaxy maps to its rotation curve through this equation.
Detailed fits \cite[(Li et al.\ 2018)]{LiRARfit} can be made with a single  
physical\footnote{One can also marginalize over the nuisance parameters of distance and inclination.} 
fit parameter: the mass-to-light ratio. There is remarkably little variation in $M_*/L_{[3.6]}$ from galaxy to galaxy, 
with the occasional, inevitable oddball: a small number of anomalous cases exist, as always happens in any large astronomical sample.

The simplicity of the organization apparent in the data has the consequence that they can only constrain a single 
fit parameter per galaxy ($M_*/L$).
Fits with traditional dark matter halo models necessarily introduce a minimum of two additional parameters per galaxy, typically
a size and mass scale. Considerable degeneracy between these parameters ensues, as is inevitable whenever three parameters
are fit to data that require only one to describe them. It is therefore impossible to uniquely constrain the properties of dark matter
halos with rotation curve fits unless strong priors from independent information are 
imposed \cite[(Katz et al.\ 2017, Li et al.\ 2019)]{Katz17,Lihalos}.
A more effective approach is to utilize the RAR: the acceleration attributable to dark matter is simply 
$\mathrm{g}_{\mathrm{DM}} = \mathrm{g}_{\mathrm{obs}} - \mathrm{g}_{\mathrm{bar}} = {\cal F}(\mathrm{g}_{\mathrm{bar}})$. 
This is not subject to multi-parameter degeneracy, being limited only by the accuracy of the stellar mass-to-light ratio
used to determine $\mathrm{g}_{\mathrm{bar}}$. In practice, a crude approximation is provided by 
a nearly constant $\mathrm{g}_{\mathrm{DM}} \approx 0.3 \times 10^{-10}\;\mathrm{m}\,\mathrm{s}^{-2}$
\cite[(Walker et al.\ 2010)]{W2010}. 

\noindent \textbf{A Common Acceleration Scale}: The three scaling relations, the BTFR, the CDR, and the RAR, are
connected by a common acceleration scale. The scale $\mathrm{g}_{\mathrm{\dagger}}$ is most obvious in the RAR, 
as it marks the transition where the
data bend away from the 1:1 line apparent at high acceleration. The same structure is also present in the CDR, which departs
from 1:1 at the surface density scale $\Sigma_{\mathrm{CDR}} \approx 860\;\mathrm{M}_{\odot}\,\mathrm{pc}^{-2}$.
Surface density is related to acceleration by Newton's constant, so these are effectively the same thing:
$\mathrm{g}_{\mathrm{CDR}} = G \Sigma_{\mathrm{CDR}}$.
Similarly, the data in the BTFR follow a line of constant acceleration $\mathrm{g}_{\mathrm{TF}} = \chi V_f^4/(G M_b)$. 
Within the uncertainties, these are the same 
scale: $\mathrm{g}_{\mathrm{\dagger}} \approx \mathrm{g}_{\mathrm{CDR}} \approx \mathrm{g}_{\mathrm{TF}}$.

The physical cause of the acceleration scale in galaxy dynamics is of fundamental importance.
There need be no such scale in galaxy dynamics at all, but it is clearly present. 
Whether this scale is unique and universal, or has some finite intrinsic scatter,
is critical to its interpretation \cite[(Di Cintio \& Lelli 2016, Desmond 2017)]{dCL16,Desomd17}.

The \textit{observed} scatter in each relation is small ($< 0.15$ dex). 
The \textit{intrinsic} scatter must be smaller as
some of the observed scatter is due to measurement errors.
An important and irreducible source of scatter is that due to variations in $M_*/L$. 
Population models suggest this should be $\ge 0.1$ dex simply from differences
in star formation histories \cite[(Bell \& de Jong 2001)]{BdJ01}. 
This implies that disk galaxies share essentially the same galaxy-averaged IMF.
It also means that any intrinsic scatter 
is small --- perhaps imperceptibly small given the inevitable scatter in $M_*/L$.
To the extent that we are able to discern, $\mathrm{g}_{\mathrm{\dagger}}$ is a fundamental scale shared by all rotating galaxies.


\end{document}